%% file: bare_conf.tex
\documentclass[conference,dvipsnames]{IEEEtran}

\IEEEoverridecommandlockouts

\usepackage{cite}

\ifCLASSINFOpdf
  \usepackage[pdftex]{graphicx}
\else
  \usepackage[dvips]{graphicx}
\fi
\usepackage{amsmath}
\usepackage{algorithmic}

\usepackage{array}

\ifCLASSOPTIONcompsoc
  \usepackage[caption=false,font=normalsize,labelfont=sf,textfont=sf]{subfig}
\else
\usepackage{subfig}
\fi
\usepackage{fixltx2e}
\usepackage{url}

\usepackage[detect-weight=true, load-configurations=binary]{siunitx}
\DeclareSIUnit{\bit}{b}
\DeclareSIUnit{\nothing}{\relax}

\usepackage{amssymb}

\usepackage{xcolor}

\usepackage{pgfplots}
\usepackage{tikz,tkz-tab}
\usepackage{circuitikz}
\usetikzlibrary{positioning, automata, graphs, trees, fit, arrows, shapes}
\usetikzlibrary{backgrounds,patterns,matrix,calc,shadows,plotmarks, circuits.logic.US}
\usetikzlibrary{decorations}

\usepackage{colortbl}

\usepackage{pgfplots}
\usepackage{pgf}
\usepackage{pgfplotstable}

\usepackage{listings}
\lstdefinelanguage{p4}
{ morekeywords={*,extern_type, attribute, type, method, extern, action, control, void, parser,state, start, transition, extract, select, default, accept, out, in, inout, return},
  sensitive=true,
  morecomment=[l]{//}, 
  morecomment=[s]{/*}{*/}, 
  morestring=[b]" 
}

\usepackage{balance}

\usepackage[bookmarks=false]{hyperref}

\newcommand\Hidden{No} 

\begin{document}
%
\title{Module-per-Object: a Human-Driven Methodology \\for C++-based High-Level Synthesis Design}


\ifthenelse{\equal{\Hidden}{No}}{
\author{\IEEEauthorblockN{Jeferson Santiago da Silva, Fran\c{c}ois-Raymond Boyer and J.M. Pierre Langlois}
\IEEEauthorblockA{Polytechnique Montr\'{e}al, Canada\\
\{jeferson.silva, francois-r.boyer, pierre.langlois\}@polymtl.ca}
\thanks{
This work was supported by the Brazilian National Council for Scientific and Technological Development - CNPq.}
}
}{
\author{\IEEEauthorblockN{First Author, Second Author and Third Author}
\IEEEauthorblockA{Authors' Institution, Authors' Country\\
Authors' email}}
}


%


\maketitle

\begin{abstract}
High-Level Synthesis (HLS) brings FPGAs to audiences previously unfamiliar to hardware design. 
However, achieving the highest Quality-of-Results (QoR) with HLS is still unattainable for most programmers. 
This requires detailed knowledge of FPGA architecture and hardware design in order to produce FPGA-friendly codes. 
Moreover, these codes are normally in conflict with best coding practices, which favor code reuse, modularity, and conciseness.

To overcome these limitations, we propose Module-per-Object (MpO), a human-driven HLS design methodology intended for both hardware designers and software developers with limited FPGA expertise. 
MpO exploits modern C++ to raise the abstraction level while improving QoR, code readability and modularity.  
To guide HLS designers, we present the five characteristics of MpO classes. 
Each characteristic exploits the power of HLS-supported modern C++ features to build C++-based hardware modules.
These characteristics lead to high-quality software descriptions and efficient hardware generation.
We also present a use case of MpO, where we use C++ as the intermediate language for FPGA-targeted code generation from P4, a packet processing domain specific language.
The MpO methodology is evaluated using three design experiments: a packet parser, a flow-based traffic manager, and a digital up-converter.
Based on experiments, we show that MpO can be comparable to hand-written VHDL code while keeping a high abstraction level, human-readable coding style and modularity. 
Compared to traditional C-based HLS design, MpO leads to more efficient circuit generation, both in terms of performance and resource utilization. 
Also, the MpO approach notably improves software quality, augmenting parameterization while eliminating the incidence of code duplication.
\end{abstract}


%
\IEEEpeerreviewmaketitle

\input{samplebody-conf}

\balance

\bibliographystyle{IEEEtran}
\bibliography{sample-bibliography} 

\end{document}

%% file: samplebody-conf.tex
\section{Introduction}\label{sec:intro}

High-level synthesis (HLS) has opened doors to an audience unfamiliar with FPGA hardware design methodology. 
Indeed, HLS tools can convert high-level and untimed C-based code into a synthesizable register-transfer level (RTL) description, a task that once had to be manually done by hardware (HW) designers. 
The RTL design flow is known to be much slower than its counterparts in software (SW) \cite{Matai:14}, since it requires a detailed description of the desired micro-architecture, including synchronization schemes, pipelining, and parallelism. 
HLS tools, on the other hand, abstract away these micro-architecture aspects allowing a faster design space exploration (DSE) through a SW development flow.

\definecolor{light-gray}{gray}{0.75}
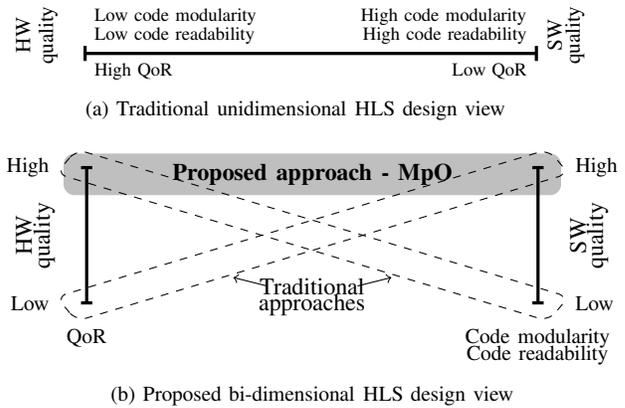
\begin{figure}[]
\subfloat[float][Traditional unidimensional HLS design view]{
\label{fig:meth:tradit}
\begin{tikzpicture}[xscale=12]
\draw[-][very thick] (0,0) -- (.5,0);

\draw [very thick] (0,-.1) -- (0,0.1);
\draw [very thick] (0.5,-.1) -- (0.5,0.1);

\node [align=center, below right] at (0,0) {\scriptsize High QoR};
\node [align=center, above right] at (0,.25) {\scriptsize Low code modularity};
\node [align=center, above right] at (0,0) {\scriptsize Low code readability};

\node [left] at (-0.05,0.25) {\rotatebox{90}{\footnotesize HW}};
\node [left] at (-0.02,0.25) {\rotatebox{90}{\footnotesize quality}};
\node [right] at (.502,0.25) {\rotatebox{90}{\footnotesize SW}};
\node [right] at (.522,0.25) {\rotatebox{90}{\footnotesize quality}};

\node [align=center, below left] at (0.5,0) {\scriptsize Low QoR};
\node [align=center, above left] at (0.5,0.25) {\scriptsize High code modularity};
\node [align=center, above left] at (0.5,0) {\scriptsize High code readability};

\end{tikzpicture}
}\\
\subfloat[float][Proposed bi-dimensional HLS design view]{
\label{fig:meth:bi_dimen}
\begin{tikzpicture}[xscale=0.75, yscale=0.45]

\draw [rounded corners, fill=light-gray, light-gray] (-.4,3.2) rectangle (8.4,4.4);
\node [align=center] at (4,3.8) {\small \textbf{Proposed approach - MpO}};

\draw[-][very thick] (0,0) -- (0,4);
\draw[very thick] (-0.1,0)  -- (0.1,0);
\draw[very thick] (-0.1,4)  -- (0.1,4);
\node [align=center, left] at (-.5,0) {\footnotesize Low};
\node [align=center, left] at (-.5,4) {\footnotesize High};
\node [align=center, below] at (0,-0.5) {\footnotesize QoR};
\node [align=center, right] at (8.5,0) {\footnotesize Low};
\node [align=center, right] at (8.5,4) {\footnotesize High};
\node [align=center, below] at (8,-.5) {\footnotesize Code modularity};
\node [align=center, below] at (8,-1) {\footnotesize Code readability};

\draw [rounded corners, dashed, rotate=26.56] (-.4,-.4) rectangle (9.35,.4);
\draw [rounded corners, dashed, rotate around={-26.56:(0,4)}] (-.4,4.4) rectangle (9.35,3.6) ;
\node [align=center, below] at (4,1) {\small Traditional};
\node [align=center, below] at (4,.55) {\small approaches};
\draw[->][] (3.2,0.5) -- (2.6,.75);
\draw[->][] (4.8,0.5) -- (5.4,.75);

\draw[-][very thick] (8,0) -- (8,4);
\draw[very thick] (7.9,0)  -- (8.1,0);
\draw[very thick] (7.9,4)  -- (8.1,4);

\node [left] at (-.8,2) {\rotatebox{90}{\small HW}};
\node [left] at (-.4,2) {\rotatebox{90}{\small quality}};
\node [right] at (8.4,2) {\rotatebox{90}{\small SW}};
\node [right] at (8.7,2) {\rotatebox{90}{\small quality}};

\end{tikzpicture}
}
\caption{HLS design approaches}
\label{fig:meth}
\end{figure}

However, achieving good Quality-of-Results (QoR) in HLS environments is sometimes unintuitive and, in some cases, not straightforward at all. 
In the HW design context, the ratio between performance and design cost normally defines the QoR standard for a given circuit. 
In FPGA design, high performance is normally associated with throughput and latency, while design cost refers to circuit area, energy consumption, and development time.


Efforts have been made to improve QoR with HLS with source-to-source transformations and code restructuring \cite{Winterstein:13,Matai:14}. 
While improving QoR, such approaches lower abstraction and make code maintenance and reuse more difficult. 
The latter two aspects are well-known problems in HLS design and they have been subject of research as well \cite{Muck:14,Muck:14_}.

Satisfactory HW QoR with HLS-based design and good SW engineering practices are often seen as incompatible \cite{hoare:1972,stroustrup:2015}. 
Indeed, the majority of HLS users are HW developers who translate RTL codes into sometimes awkward HW-oriented C-based descriptions. They attempt to reproduce RTL-level microarchitectural expressiveness while still accelerating the FPGA design cycle through HLS design flow. 
Such HW-oriented C descriptions lead to incomprehensible codes difficult to reuse by other designers.

Although existing HLS approaches can sometimes deliver good code readability and modularity, and still produce good results, this is most often not the case. Normally, HLS development trades-off HW QoR and SW quality, following a sort of unidimensional view, as illustrated in Fig.~\ref{fig:meth:tradit}.
However, a bi-dimensional HLS approach is required. Indeed, a bi-dimensional perspective highlights independence between HW QoR and SW quality. 
Fig.~\ref{fig:meth:bi_dimen} shows the design space of this novel bi-dimensional HLS view. 
In fact, in the course of this work, we show that using our approach, it is possible to increase HW QoR and SW quality simultaneously by employing modern and high-quality C++ constructs, which leads to cleaner codes and reduces duplication.

In this context, we present design guidelines for C++-based HLS design targeting both HW \textit{and} SW designers. 
We present several C++ high-level constructs and, whenever possible, we show their correspondence in the generated HW. 
The HLS methodology we propose is called Module-per-Object (MpO). It is meant to be human-driven and used by ordinary programmers with limited HW expertise, not only by FPGA experts. 
We aim to close the gap between QoR and code modularity and readability. 
We use the results obtained by traditional HLS design as HW QoR metric.
We focus on code modularity and readability as SW quality metrics. Code modularity is evaluated by the capability of reuse of a given module while code readability is related to the code expressiveness and conciseness.

As a final goal, we intend to widen FPGA usage by SW programmers by raising the FPGA development abstraction. 
Indeed, higher design abstractions allow programmers to use a single version of their code to run on an \texttt{x86} CPU or be synthesized for an FPGA device \cite{Cong:2011}. 
To do so, we propose to exploit high-level modern constructs and the Standard Template Library (STL). Such constructs are well known by SW developers to improve code readability \cite{stroustrup:2015}. 
We target QoR and code readability and modularity by extensively employing templated classes and structures that can tune the C++ objects according to design needs. 
In addition, we discuss the possibility of adopting templated C++ classes as an intermediate language to be used alongside a Domain Specific Language (DSL). 
The main contributions of this work are as follows: 

\begin{itemize}
\item A methodology called Module-per-Object, a design pattern for HLS design that simultaneously achieves high modularity, readability, and QoR (\S~\ref{sec:module_object});
\item The extensive use of synthesizable templated C++ data structures and constructs to improve QoR and modularity with HLS (\S~\ref{sec:module_object});
\item A case-study on using C++ as an intermediate language for automatic code generation of a packet parser written in the P4 language targeting FPGAs (\S~\ref{sec:p4_pipe});
\item Based on three specific use-cases, we have identified HLS tools deficiencies that prevent exploiting the full capabilities of high-level constructs, and we propose guidance for HLS designers and hints for future HLS tool releases (\S~\ref{sec:limitations}); and
\item An evaluation of the benefits brought by the MpO approach on three design examples: a packet parser, a flow-based traffic manager, and a digital-up converter (\S~\ref{sec:results}). 
\end{itemize}


\begin{table*}[]
\centering
\caption{Summary of C++ features used in this work}
\label{tab:Cpp_for_synth}
\small{
\begin{tabular}{lll}
\hline
\textbf{Constructs}               & \textbf{Benefits}                                                                                                         & \textbf{Version}                                  \\\hline
\rowcolor[HTML]{EAEAEA} 
Fixed-point types        & Fixed-point arithmetic                                                                                           & C++98, vendor dependent  \\
(Variadic) Templates                & Parameterizable design                                                                                           & (C++11), C++98                                    \\
\rowcolor[HTML]{EAEAEA} 
Classes                  & OO paradigm, encapsulation, inheritance, polymorphism & C++98                                    \\
Template metaprogramming & Compile-time calculation, performance improvement & C++98                                    \\
\rowcolor[HTML]{EAEAEA} 
STL                      & Modularity, code reuse, standardization                                                                          & \textgreater~C++98, in constant evolution\\
Data containers          & Data storage and encapsulation                                                                                   & \textgreater~C++98, in constant evolution                                    \\
\rowcolor[HTML]{EAEAEA} 
Algorithms               & Standardization, code reuse                                                                                      & \textgreater~C++98, in constant evolution\\
Iterators, range-based \texttt{for} loops & Syntax sugaring, easier container iteration                                                                             & C++11                                    \\
\rowcolor[HTML]{EAEAEA} 
Lambda expressions       & Function pointer properties                                                                                    & C++11 \\
\texttt{constexpr} variables and functions       & Compile-time calculation, performance improvement & C++11 \\
\rowcolor[HTML]{EAEAEA} 
\texttt{auto}, \texttt{decltype}       & Automatic type inference                                                                                    & C++11 \\\hline                                  
\end{tabular}}
\end{table*}

\section{Related Work}\label{sec:related_works}


\subsection{QoR Improvements in HLS-based Design}\label{sec:related_works:qor}



Liang \textit{et al.} \cite{liang:12} conducted a study on how to restructure C codes in order to improve QoR with HLS for several different benchmarks. Their results showed up to $126\times$ performance improvement over a pure software implementation, which were obtained after various rounds of code refactoring and \texttt{\#pragma} insertions, which requires extensive HW expertise. In addition, when comparing to hand-crafted RTL design, their results are up to $20\times$ worse. Also, the authors affirm that, in some cases, improving QoR conflicts with good SW engineering practices. Matai \textit{et al.} \cite{Matai:14} presented a methodology for code restructuring with HLS targeting FPGA devices. However, the transformed codes are unintuitive and not portable. Similar research was conducted by Homsirikamol and Gaj \cite{Homsirikamol:14} and Liu \textit{et al.} \cite{Liu:2016}. Zhou \textit{et al.} have presented Rosetta \cite{Zhou:2018}. Rosetta is a benchmark suite for HLS-driven FPGA design. The benchmarks have been meticulously coded and tuned for state-of-the-art HLS tools. While such practices improve performance and reduce FPGA area, in most cases, the source code is unreadable for a non-FPGA expert.

Source-to-source transformations have been explored by Winterstein \textit{et al.} \cite{Winterstein:13,Winterstein:14}. The authors have proposed a framework that performs source-to-source transformations on the original C code in order to ensure synthesizability. The authors claim that the produced code is human-readable. Automated source-to-source transformations can result in descriptions that might not exactly match the original code. Gao \textit{et al.} \cite{Gao:2016} and Cong \textit{et al.} \cite{Cong:17} have done similar research.

\subsection{Raising the Abstraction Level in HLS}\label{sec:related_works:hll}

Cong \textit{et al.} \cite{Cong:2011} have conducted a thorough study on HLS methods and tools. They have as well evaluated the performance of the former AutoESL's HLS tool. The authors have presented a design methodology for HLS-driven FPGA design, which includes code reusing practices through C++ templates.

Muck and Frohlich \cite{Muck:14,Muck:14_} have exploited advanced and metaprogrammed C++ constructs to create compatible codes for both CPUs and FPGA devices. The authors present guidelines for FPGA-friendly pointer handling and static polymorphism implementation\cite{Coplien:1995}. According to the authors, the resulting overhead in having reusable and modular unified C++ codes is worthwhile. The area and performance overhead are up to $30\%$ and $50\%$, respectively, compared to HW-oriented C++ design. Our work leverages their ideas by employing several other C++11 constructs and by comparing the achievable results with RTL implementations.

Thomas \cite{Thomas:2016} has presented a DSL library targeting recursion with C++ HLS tools described using C++11 constructs. The author has shown how compile-time metaprogramming and lambda expressions can leverage HLS-driven HW design. Indeed, in our work, we have confirmed that such constructs \textit{can} be used by HLS designers, eventually leading to higher QoR, while raising the  abstraction. 
Similar research was conducted by Richmond \textit{et al.} \cite{Richmond:2018}. Recently, Eran \textit{et al.} \cite{Eran:19} have proposed HLS-friendly design patterns for packet processing exploiting the capabilities of modern post C+11.

Zhao and Hoe \cite{Zhao:17} have assessed HLS-based flow in structural design. Their results for a network-on-chip implementation are comparable with a self-generated RTL approach. The area and performance results vary according to the network topology, ranging from $+1\%$$\sim$$+23\%$ in lookup tables (LUTs), $-71\%$$\sim$$-54\%$ in flip-flops (FFs), and $-14\%$$\sim$$+24\%$ in clock frequency. Their approach does not explore in depth the capabilities of C++ constructs supported by the HLS tool, which improves code modularity and readability.

Oezkan \textit{et al.} \cite{Oezkan:17} have also exploited templated C++ classes to build an image processing library targeting FPGA devices. The authors make extensive use of templates to generate highly parameterizable C++ classes. One of their final remarks is that the more the code is written in a ``hardware design manner", the better its synthesis is. This ``hardware manner" coding style lowers the abstraction, which could be alleviated by exploiting the potential of the available high-level constructs of the STL, augmenting thus code readability, avoiding code duplication, and improving code maintenance.



\subsection{HLLs as Intermediate Representation in FPGA Design}\label{sec:related_works:ir}


Other researchers have pointed to the use of DSLs for FPGA design \cite{Kapre:16}. Although increasing the development abstraction, such languages need to be converted into synthesizable RTL code, a process similar to what is done by HLS tools. Examples of such DSLs can be found in most varied domains, ranging from signal/image processing to network applications. 

In the network domain, several works have used HLLs, such as P4 \cite{Bosshart:14}, for FPGA implementation. P4FPGA \cite{Wang:2017} is a framework for fast prototyping of network functions described in P4. P4FPGA uses BlueSpec Verilog as intermediate representation idiom, which requires a proprietary compiler to generate synthesizable RTL. The approach proposed by Khan \cite{Khan:2017} uses off-the-shelf HLS tools, however, it is difficult to evaluate the real impact of this work due to the lack of details provided. While Emu \cite{Sultana:2017} is not used alongside a higher level network DSL, it could have been, since it comprises a set of standard network libraries written in C\# in an object-oriented fashion that are compiled to Verilog using Kiwi \cite{Singh:2008}. These approach is similar to what Silva \textit{et al.} \cite{SantiagodaSilva:2018} have done for a P4-compatible packet parser.

%

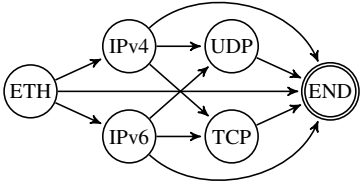
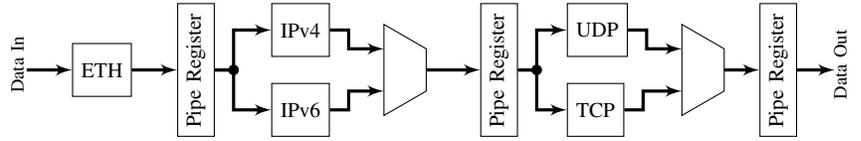
\begin{figure*}[t]
\subfloat[float][Example of a parser graph]
{
\begin{tikzpicture}[->,>=stealth',shorten >=1.25pt,auto,node distance=.75cm, semithick, every fit/.style={rectangle, draw, inner sep=1.25pt}] \tikzstyle{every state}=[minimum size=2em, inner sep=1.25pt]
  \node[state]	    (Eth)  { \footnotesize ETH};
  \node[state,anchor=center]		(IPv4) at ($(Eth.east) + (.95cm ,.6cm)$) { \footnotesize IPv4};
  \node[state,anchor=center]		(IPv6) at ($(Eth.east) + (.95cm ,-.6cm)$) { \footnotesize IPv6};
  \node[state,anchor=center]		(UDP) at ($(IPv4.east) + (1.0cm ,.0cm)$) {\footnotesize UDP};
  \node[state,anchor=center]		(TCP) at ($(IPv6.east) + (1.0cm ,.0cm)$) {\footnotesize TCP};
  \node[state, accepting,anchor=center] (End) at ($(TCP.east) + (.95cm ,.6cm)$) {\footnotesize END};
    
  \path (Eth)  edge (IPv4)
          (Eth)  edge (IPv6)
          (Eth)  edge (End)
		  (IPv4) edge (UDP)          
		  (IPv4) edge (TCP)
		  (IPv4) edge [bend left=55] (End)
		  (IPv6) edge (UDP)          
		  (IPv6) edge (TCP)
		  (IPv6) edge[bend right=55] (End)
		  (UDP) edge (End)
		  (TCP) edge (End);
\end{tikzpicture}
\label{fig:graph:graph}
}\qquad
\subfloat[][Pipeline organization for the example packet parser]
{

\tikzset{%
  input/.style    = {coordinate}, 
  output/.style   = {coordinate} 
}
\tikzstyle{header} = [draw, rectangle, 
    minimum height=2.0em, minimum width=2.0em]
\tikzstyle{pipe} = [draw, rectangle, 
    minimum height= 0.4em, minimum width=2.0em]
\tikzstyle{mux} = [ trapezium,   draw,   
                    shape border rotate = 270,
  inner xsep=8pt, inner ysep=8pt, outer sep=0pt,
  minimum height=.1cm]
\tikzstyle{input} = [coordinate]
\tikzstyle{output} = [coordinate]

\begin{tikzpicture}[auto,>=latex',distance=1cm]
    \node [input, name=input, anchor=center] at (0cm, 0cm){};
    \node [header, right of=input, align=center] (ETH) {\footnotesize ETH};  
    \node [pipe, right of=ETH, node distance=1.25cm, rotate=90] (pipea) {\footnotesize Pipe \footnotesize Register};
    \node [header, right = .75cm of pipea.south east, anchor = north west, align=center] (IPv4) {\footnotesize IPv4};
    \node [header, right = .75cm of pipea.south west,anchor=south west,  align=center] (IPv6) {\footnotesize IPv6};
    \node [mux, anchor=center] (mux) at ($(IPv4.south east |- pipea.south) + (1.0cm ,0cm)$) {}; 
    \node [pipe, right of=mux, node distance=1.25cm, rotate=90] (pipec) {\footnotesize Pipe  \footnotesize Register};
    \node [header, right = .65cm of pipec.south east, anchor = north west, align=center] (UDP) {\footnotesize UDP};
    \node [header, right = .65cm of pipec.south west,anchor=south west,  align=center] (TCP) {\footnotesize TCP};
    \node [mux, anchor=center] (mux1) at ($(UDP.south east |- pipec.south) + (1.0cm ,0cm)$) {}; 
    \node [pipe, right of=mux1, node distance=1cm, rotate=90] (piped) {\footnotesize Pipe \footnotesize Register};
    \node [output, right of=piped, node distance=.75cm] (output) {};
    
    \draw [draw,->, very thick] (input) -- node[label={[xshift=-0.2cm, yshift=-0.2cm, rotate=90]\scriptsize Data \scriptsize In}]{} (ETH);
    \draw [->, very thick] (ETH) -- node {} (pipea);
    \draw [->, very thick] (pipea) --+(0.5cm,0) |- node {} (IPv4.west);
    \draw [->, very thick] (pipea) --+(0.5cm,0) |- node {} (IPv6.west);
    \fill ($(pipea)+(0.5cm,-0cm)$) circle (2pt);
    \draw [->, very thick] (IPv4.east) --+(0.25cm,0)|- node {} (mux.north west);
    \draw [->, very thick] (IPv6.east) --+(0.25cm,0)|- node {} (mux.south west);
    \draw [->, very thick] (mux) -- node  {} (pipec);
    \draw [->, very thick] (pipec) --+(0.5cm,0) |- node {} (UDP.west);
    \draw [->, very thick] (pipec) --+(0.5cm,0) |- node {} (TCP.west);
    \fill ($(pipec)+(0.5cm,-0cm)$) circle (2pt);
    \draw [->, very thick] (UDP.east) --+(0.25cm,0)|- node {} (mux1.north west);
    \draw [->, very thick] (TCP.east) --+(0.25cm,0)|- node {} (mux1.south west);
    \draw [->, very thick] (mux1) -- node  {} (piped); 
    \draw [draw,->, very thick] (piped) -- node[label={[xshift=0.55cm, yshift=-0.2cm, rotate=90]\scriptsize Data \scriptsize Out}]{} (output);

\end{tikzpicture}
\label{fig:graph:final_pipe}
}\hfill
\caption{Representation of a packet parser}
\label{fig:graph}
\end{figure*}

\section{MpO HLS Methodology}\label{sec:module_object}


\subsection{Overview of the MpO Methodology}

We propose the Module-per-Object (MpO) HLS methodology, in which we define the concept of \textit{``module''} as a C++ object that logically represents a self-contained functional unit. To do so, this work exploits high-level constructs available in C++11, and that are supported by Xilinx Vivado HLS, to improve QoR while keeping a very high level of abstraction. 
Inspired by Cong \textit{et al.} \cite{Cong:2011}, Table~\ref{tab:Cpp_for_synth} summarizes the synthesizable C++ constructs used in this work.

To increase code modularity and readability, our approach uses the concept of an MpO base class, which abstracts common functionalities between different modules. Consequently, this approach allows to reuse the same source code to describe functional modules with similar behavior. The five characteristics of an MpO class are: 1) Templates: class parameterization and code modularity (\S~\ref{sec:templates}); 2) Systematic utilization of \texttt{const} and \texttt{constexpr} variables for static objects (\S~\ref{sec:constexpr}); 3) STL constructs: zero-overhead abstraction, code reuse and modularization (\S~\ref{sec:STL}); 4) Inheritance and static polymorphism (when appropriate): code reuse and modularization (\S~\ref{sec:polymorphism}); and, 5) Smart constructors: constant class member initialization (\S~\ref{sec:constructors}).

The main idea is to write generic code that is specialized at compile time. Generics codes, exploiting templates (1), STL constructs (3), and inheritance and static polymorphism (5), allow writing more compact and reusable code, reducing code duplication. Specialized objects also help reducing resource usage by allowing specific pieces of hardware to be precisely inferred. Indeed, \texttt{const} and \texttt{constexpr} variables (2) give hints to the compiler to perform constants propagation that can be used in conjunction with smart constructors (5) for class member initialization.



\subsection{Illustrative Use Case: a Packet Parser}\label{sec:case_study}


We demonstrate the viability of the proposed methodology with the design of a packet parser as a use case. A packet parser determines the set of valid protocols supported by a network device and extracts the required header fields that are to be matched in the packet processing pipeline.

A packet parser can be modeled at a high-level with a directed acyclic graph (DAG), where nodes represent protocols and edges are protocol transitions \cite{Gibb:2013}. A parser is implemented as an abstract state machine (ASM), performing state transition evaluations at each parser state. States belonging to the path connecting the first state to the last state in the ASM compose the set of supported protocols of a network equipment. A packet-processing language, such as P4 \cite{Bosshart:14}, can be used to describe such an ASM. 
Details on the implementation of a packet parser in FPGA can be found in \cite{SantiagodaSilva:2018}. Fig.~\ref{fig:graph:graph} illustrates a parser graph for a layer-4 network device while Fig.~\ref{fig:graph:final_pipe} shows its possible hardware realization.
%
%



\subsection{Specializing Classes with Templates}\label{sec:templates}

Templates are fundamental to correctly parameterize an MpO base class. Indeed, class templates allow generic code to be fine-tuned for different design instances, favoring code reuse, reducing duplication while generating results comparable to hand-tuned codes. 

Referring to Fig.~\ref{fig:graph:graph}, the nodes of the parser graph share common properties and may share the same code, being a great starting point for an MpO base class. Listing~\ref{list:header} presents an example of an MpO base class that describes a node of the parser. For simplicity, only relevant code fragments are shown and cannot be compiled as is.


\begin{lstlisting}[float,label=list:header,caption=The \texttt{Header} MpO C++ base class.]
template<(*@{$\cdots$}@*), class T_HeaderLayout, class T_DHeader> (*@{\label{list:header:template1}}@*)(*@{\label{list:header:template2}}@*)
class Header {
	protected:
		typedef ap_uint<numbits(B2b(N_Size))> RXBitsType;(*@{\label{list:header:RXBitsType}}@*)
		const T_HeaderLayout HeaderLayout; (*@{\label{list:header:HeaderLayout}}@*)
		const ShiftType stateTransShiftVal;
		const array<bool, ARR_SIZE> HeaderBusCompVal;
		RXBitsType rxBits;
	public:
		template<typename T, typename F> (*@{\label{list:header:callable}}@*)
		const T init_array(const F& func) const {
			typename remove_cv<T>::type arr {};
			for (auto i = 0; i < arr.size(); ++i)
				arr[i] = func(i);
			return arr;
		}
		Header (const headerIDType instance_id, const T_HeaderLayout& HLayout) :(*@{\label{list:header:constructor}}@*)   
			(*@{$\cdots$}@*)
			HeaderLayout(HLayout),
			stateTransShiftVal{shift_def(B2b(N_Size), N_BusSize,(*@{\label{list:header:stateTransShiftVal}}@*) 
				(HLayout.KeyLocation.first + HLayout.KeyLocation.second))
			},
			HeaderBusCompVal( (*@{\label{list:header:callable_invokation}}@*)
				init_array<decltype(HeaderBusCompVal)>(
					[HLayout](size_t i) { 
						return (HLayout.ArrLenLookup[i] >> numbits(N_BusSize)) > 0;
					}
				)
			),
		{ (*@{$\cdots$}@*) } // end of constructor
		void StateTransition(const PktDataType& PktIn);
		void PipelineAdjust((*@{$\cdots$}@*));(*@{\label{list:header:PipelineAdjust}}@*)
		void HeaderAnalysis(const PktDataType& PktIn,PHVDataType& PHV,PktDataType& PktOut);
};
\end{lstlisting}

The class presented in Listing~\ref{list:header} is parameterized with four template parameters (line \ref{list:header:template1}). The two first parameters, omitted in the listing, are integers and they are used to configure the arbitrary-sized integers. \texttt{T\_HeaderLayout} is a struct type derived from a template. This type is used to declare the class member \texttt{HeaderLayout} on line~\ref{list:header:HeaderLayout}, which represents the expected header layout to be processed. The last template parameter, \texttt{T\_DHeader}, is also a type. However, this type is used to allow static polymorphism of methods of the \texttt{Header} class; therefore, it represents a type that is derived from the \texttt{Header} class itself \cite{Coplien:1995}.

Consequently, with the extensive use of templates, an MpO base class provides a high-degree of configurability to MpO class objects. Thus, MpO base classes contribute to more reusable and compact code. The graph described in Fig.~\ref{fig:graph:graph} is an example where node is a different C++ object, sharing the same source code, described in Listing~\ref{list:header}, using different template parameters.



\subsection{Specializing Operands with \texttt{constexpr}}\label{sec:constexpr}


In MpO, we use \texttt{constexpr} functions and variables to set accurate bus sizes in a generic fashion, which leads to faster and more compact circuits while configurable yet synthesizable C++ descriptions are used. Also, \texttt{constexpr} functions are more comprehensible compared to the their equivalents using older C++ versions. Indeed, they allow template specialization and alleviate a task that before C++11 was only possible through template metaprogramming and partial template specialization.


The type \texttt{RXBitsType} in Listing~\ref{list:header} line~\ref{list:header:RXBitsType} is such an example. The functions \texttt{numbits()}, \texttt{B2b()}, and \texttt{shift\_def()} in Listing~\ref{list:header} are examples of \texttt{constexpr} functions. In \cite{cpp:constexpr}, we present the implementation of the \texttt{numbits()} function along its verbose equivalent described in C++03. This function returns the size in bits to represent an arbitrary-sized integer.



One can benefit of compilers' ability to propagate constants by using \texttt{constexpr} functions to initialize class members in constructors. An example is the protected member \texttt{stateTransShiftVal} of the \texttt{Header} class in Listing~\ref{list:header}~line~\ref{list:header:stateTransShiftVal}, whose value is compile-time resolved when the class constructor is called (line~\ref{list:header:constructor}), becoming a hardwired value in the HW implementation.

\subsection{Exploiting STL Constructs}\label{sec:STL}

\begin{lstlisting}[float,label=list:state,caption=The \texttt{StateTransition()} method]
template<(*@{$\cdots$}@*)> void Header<(*@{$\cdots$}@*)>::StateTransition(const PktDataType& PktIn){
	typedef decltype(HeaderLayout.Key.front().KeyVal) KeyType;(*@{\label{list:header:KeyType}}@*)
	const KeyType DataInMask = createMask(HeaderLayout.KeyLocation.second);
	KeyType packetKeyVal = (PktIn.Data >> stateTransShiftVal) & DataInMask;
	if (!NextHeaderValid && (rxBits > HeaderLayout.KeyLocation.first))
		for (auto key : HeaderLayout.Key)(*@{\label{list:header:for}}@*)
			if (key.KeyVal == (packetKeyVal&key.KeyMask)) {
				NextHeader = key.NextHeader;
				NextHeaderValid = true;
			}
}
\end{lstlisting}

STL constructs raise the development abstraction and ease code readability and maintenance, characteristics favored by the MpO methodology. 

Listing~\ref{list:state} shows how such constructs can be used to describe a possible implementation of a state evaluation function in a parser ASM. Its goal is to search the incoming packet stream to determine if there is a valid protocol transition for a given ASM state. To do so, the members \texttt{Key} and \texttt{KeyLocation} of the \texttt{HeaderLayout} \texttt{struct} are used. \texttt{Key} is an STL \texttt{array} container, composed of another data structure that holds information regarding the value to be matched and which is the next header transition in case of a match. \texttt{KeyLocation} is an STL \texttt{pair} type, where the first member is the key location in the incoming data stream and the second member is the key size in bits.

An \texttt{array<Type, Size> Array} container is a fixed-sized array similar to the array declaration \texttt{Type Array[Size]} in ISO C. However, since it belongs to the STL, it includes some useful built-in methods, such as \texttt{size()} and \texttt{front()}. These method calls can be resolved at compile time, and therefore they can be used to parameterize types and to set fixed loop bounds. One example of such utilization is shown in Listing~\ref{list:state} in the \texttt{KeyType} type definition on line~\ref{list:header:KeyType}. To define this new type, we use the \texttt{decltype} keyword. Again, one \texttt{constexpr} function is used, \texttt{createMask()}, to allow constant propagation on variable \texttt{DataInMask}.

Also, STL \texttt{array}s, such as the \texttt{HeaderLayout.Key}, allow the use of iterators in a range-based \texttt{for} loop to iterate over the array. Such constructs lead to safer and more compact code since it is not required to calculate the iteration indexes or to specify loop bounds. Such an example is the \texttt{for} statement shown in Listing~\ref{list:state} line~\ref{list:header:for}. In addition, automatic type resolution can be used with the \texttt{auto} keyword to determine the type of the loop iterator, simplifying the code as well.

According to our experiments, using STL constructs did not introduce overhead in terms of QoR. However, the increased code readability and modularity is noticeable, specially when dealing with data containers, such as \texttt{array}, by minimizing the need for raw pointer manipulation as required in C~\cite{cpp:zero_abstraction}.

\subsection{Inheritance and Static Polymorphism}\label{sec:polymorphism}

\begin{lstlisting}[float,label=list:header_layout,caption=Example of static polymorphism]
template<(*@{$\cdots$}@*), class T_DHeaderFormat>class HeaderFormat {
	ap_uint<HSIZE_BITS> getHeaderSize(const ap_uint<HSIZE_BITS>& expr_val) const  (*@{\label{list:header_layout:getHeaderSize}}@*) 
		{ return static_cast<T_DHeaderFormat*>(this)->getSpecHeaderSize(expr_val); }
};
template<(*@{$\cdots$}@*)> (*@{\label{list:header_layout:crtp1}}@*)
class varHeaderFormat : public HeaderFormat< (*@{$\cdots$}@*) , const varHeaderFormat< (*@{$\cdots$}@*) >> { (*@{\label{list:header_layout:crtp2}}@*)
	ap_uint<HSIZE_BITS> getSpecHeaderSize(const ap_uint<4>& ihl) const  (*@{\label{list:header_layout:getSpecHeaderSize}}@*)
		{ return ((0x4*ihl)*0x8); }(*@{\label{list:header_layout:size}}@*)
};
\end{lstlisting}

The MpO methodology favors code reuse by employing inheritance whenever possible. Inheritance greatly improves code modularity and maintainability by reducing code replication. MpO exploits C++ to leverage the DRY (don't repeat yourself) design guideline. Indeed, C++ offers adequate artefacts for improving inheritance, such as polymorphic methods and virtual classes.


Virtual classes are, to date, not supported by HLS vendors. However, inheritance and static polymorphism are allowed.

For the packet parser, it is of interest to keep the same method calls even if variable- and fixed-size headers are processed in a different manner. To do so, static polymorphism is a C++ mechanism that can be used with MpO.

To parse fixed-sized headers, all needed information is known at compilation time. When processing variable-sized headers, the header length must be retrieved from the header information itself. To do so, the \texttt{T\_HeaderLayout} type in Listing~\ref{list:header} implements static polymorphism to retrieve both fixed- or variable-sized header length information using the same method call. The \texttt{T\_HeaderLayout} definition is shown in Listing~\ref{list:header_layout}.


In Listing~\ref{list:header_layout}, the \texttt{HeaderFormat} is the base \texttt{struct}. The \texttt{struct} \texttt{varHeaderFormat} and \texttt{fixedHeaderFormat} (not shown in the code extract) are derived from \texttt{HeaderFormat}. Note that to allow static polymorphism, we use the Curiously Recurring Template Pattern (CRTP) technique \cite{Coplien:1995} as in \cite{Muck:14,Muck:14_}, where the derived class is passed as a template parameter to the base class (lines~\ref{list:header_layout:crtp1}-\ref{list:header_layout:crtp2}). By doing so, the compiler is able to statically resolve pointer conversions, which results in a synthesizable description. In this example, the implementation of the \texttt{getHeaderSize()} method (line~\ref{list:header_layout:getHeaderSize}) is done in the derived struct (line~\ref{list:header_layout:getSpecHeaderSize}).

The base class \texttt{Header} from Listing~\ref{list:header} also supports CRTP to implement static polymorphism. Two classes are derived from the \texttt{Header} class: the \texttt{FixedHeader} class and the \texttt{VariableHeader} class. Similarly to what is done with the \texttt{HeaderFormat} from Listing~\ref{list:header_layout}, the classes derived from the \texttt{Header} class have their own implementation for the method \texttt{PipelineAdjust()} (Listing~\ref{list:header} line~\ref{list:header:PipelineAdjust}). This method is responsible for keeping the output data bus aligned for the next processed header. To process fixed-sized headers, fixed bit-shift operations suffice for this alignment while barrel-shifters are required when dealing with variable-sized headers.


A naive barrel-shifter implementation in FPGAs is based on a chain of multiplexers, which results in $O(N log(N))$ area complexity and $O(log(N))$ delay. Contrary to ASIC design, implementing wide multiplexers can be costly in FPGAs, having normally the same complexity as an adder \cite{Cong:2011}. Thus, avoiding wide multiplexers is desired when designing efficient FPGA HW. In the parser, the number of bits to be shifted is a function of the current header size. Once we are dealing with wide data buses and the size of the processed headers is well-constrained by a formula (Listing~\ref{list:header_layout}~line~\ref{list:header_layout:size}) in which only a few set of values are valid, then a natural choice is to use a small lookup table storing only the set of valid shift operands.

\subsection{Smart Constructors}\label{sec:constructors}

Class constructors can be used to initialize constant class members, which leads to more efficient circuits, as in the constant lookup table of shift values in the previous section.

An example of a smart constructor that makes use of a templated function is shown in Listing~\ref{list:header}~line~\ref{list:header:callable}. 
The function is called in the constructor in line~\ref{list:header:callable_invokation} to initialize the \texttt{const} class member \texttt{HeaderBusCompVal}. 
Note that the templated function uses a lambda expression as a callable parameter.

In C++, templated functions and objects allow callable objects to be passed as parameters to functions. Callable parameters allow functions to be reused, thus reducing code duplication.
Such callable parameters can be function pointers, functors (function objects), or lambda expressions, which were introduced in C++11. 
Functors are objects with a single method, which once constructed can be called as a function. 
Modern compilers have the ability to optimize the object construction, inlining the code within the scope it is called. 
More interestingly, lambdas are local functions which are stored as variables, while allowing parameter passage and context capturing. 
In fact, lambdas are syntax sugaring for functors \cite{Thomas:2016}. 
Indeed, for the same functionality, both the functor and the lambda implementation generate the same assembly (and LLVM) code \cite{cpp:lambda}.

Function pointers are unsynthesizable constructs by most HLS tools. Thus, functors and lambdas are alternative yet synthesizable ways to emulate function pointers. Besides being convenient and elegant, lambdas can contribute to more efficient HW generation by enabling constant propagation when initializing constant class members in class constructors.

\section{Packet-parser Generation from P4}\label{sec:p4}


\subsection{Top-Level Pipeline}\label{sec:p4_pipe}

Until now, we have described how a single HW module can be described using the proposed MpO approach. Several instances of the generic \texttt{Header} class from Listing~\ref{list:header} can be specialized to generate different HW modules. Therefore, the proposed MpO methodology from \S~\ref{sec:module_object} can be used to implement a complete packet parser.

Listing~\ref{list:parser} shows a possible implementation for the packet parser illustrated in Fig.~\ref{fig:graph}. The code in this listing is automatically generated from a P4 description \cite{Bosshart:14}. Details on the internal parser micro-architecture and the optimization steps for code generation are subject of previous work \ifthenelse{\equal{\Hidden}{Yes}}{[hidden for blind review]}{\cite{SantiagodaSilva:2018}}.

The generated HW architecture from Listing~\ref{list:parser} is in accordance to the parser pipeline organization shown in Fig.~\ref{fig:graph:final_pipe}. This is ensured by the \texttt{static} declaration of the parser node objects (line~\ref{list:parser:FixedHeader} and \ref{list:parser:VariableHeader}), in a similar approach to what Zhao and Hoe have proposed \cite{Zhao:17}. The \texttt{static} keyword is used to declare stateful header objects. The pipeline is therefore inferred according to the data dependency graph. Conditional inputs in a given pipeline stage or in the output are resolved with the ternary (\texttt{?:}) C operator (line~\ref{list:parser:tmpPIn_3}, \ref{list:parser:tmpPIn_4}, and \ref{list:parser:PktOut}), which generates a multiplexer in the final HW \cite{Kim:2017}. 


\subsection{Adapting MpO to Current HLS tools}\label{sec:limitations}

Vivado HLS supports C, C++, and SystemC for synthesis and simulation. The most recent C++ version supported by Vivado HLS dates from 2011. However, Vivado HLS does not fully support this C++ version, limiting the spectrum of standard high-level constructs that can be used to raise the development abstraction.


This work makes extensive use of the C++11 STL. While some constructs available in the library are expected to fail during synthesis, such as lists and maps, fixed-bounded constructs are well supported. These constructs, such as the standard \texttt{array} and \texttt{pair}, are described as classes in the STL and their operators are defined as functors in these classes. During synthesis, when facing each of these operators, Vivado HLS performs automatic function inlining for the method describing one operator, which leads to longer synthesis time and memory usage. The decision to use these STL constructs is, therefore, a trade-off between the synthesis time and the flexibility provided by these constructs.

During this work, we have struggled to correctly implement dynamic polymorphism with Vivado HLS.  Static polymorphism through CRTP was the only found solution for polymorphism in this work. However, even static polymorphism is limited. Derived classes can access neither local members nor base class members. Such data accesses cause an invalid pointer reinterpretation error under synthesis. The detour for such errors is to pass the necessary operands as function parameters to the callee methods in the derived class. Accessing static members in the base class does not cause any error.

Modern compilers are able to devirtualize virtual methods of dynamically polymorphic classes at compile time and to inline the code in derived classes. Clang, for instance, is capable of devirtualizing with the compiler optimization flag set to \texttt{-O2}~\cite{cpp:optim}. 
 However, Vivado HLS does not support the compiler optimization flags. Since the optimization flag has no effect on Vivado HLS, and its own synthesis pass is not able to infer the virtual type, dynamic polymorphism cannot be synthesized. Thus, as already concluded by other researchers \cite{Noronha:2017}, borrowing some front-end optimization techniques from modern compilers may be useful in the HLS world. 



\begin{lstlisting}[float,label=list:parser,caption=The Parser pipeline]
void Parser(const PktDataType& PktIn, EthPHVDataType& eth_PHV, (*@{$\cdots$}@*) , PktDataType& PktOut) { 
	array<PktDataType, 5> tmpPIn, tmpPOut;
	static FixedHeader<(*@{$\cdots$}@*)> eth ((*@{$\cdots$}@*)); (*@{\label{list:parser:FixedHeader}}@*)
	static EthPHVDataType tmpEthPHV;
	static VariableHeader<(*@{$\cdots$}@*)> ipv4((*@{$\cdots$}@*)); (*@{\label{list:parser:VariableHeader}}@*)
	static Ipv4PHVDataType tmpIpv4PHV;
	(*@{$\cdots$}@*)
	tmpPIn[0] = PktIn;
	eth.HeaderAnalysis(tmpPIn[0],tmpEthPHV,tmpPOut[0]);
	eth_PHV = tmp_eth_PHV;
	tmpPIn[1] = tmpPOut[0];
	ipv4.HeaderAnalysis(tmpPIn[1],tmpIpv4PHV,tmpPOut[1]);
	ipv4_PHV = tmpIpv4PHV;
	tmpPIn[2] = tmpPOut[0];
	ipv6.HeaderAnalysis(tmpPIn[2],tmpIpv6PHV,tmpPOut[2]);
	ipv6_PHV = tmpIpv6PHV;
	tmpPIn[3] = (tmpIpv4PHV.Valid)?tmpPOut[1]:tmpPOut[2];(*@{\label{list:parser:tmpPIn_3}}@*)
	udp.HeaderAnalysis(tmpPIn[3],tmpUdpPHV,tmpPOut[3]);
	udp_PHV = tmpUdpPHV;
	tmpPIn[4] = (tmpIpv4PHV.Valid)?tmpPOut[1]:tmpPOut[2];(*@{\label{list:parser:tmpPIn_4}}@*)
	tcp.HeaderAnalysis(tmpPIn[4], tmpTcpPHV, tmpPOut[4]);
	tcp_PHV = tmpTcpPHV;
	PktOut = (tmpUdpPHV.Valid)?tmpPOut[3]:tmpPOut[4];(*@{\label{list:parser:PktOut}}@*)
}
\end{lstlisting}

\section{Experimental Results}\label{sec:results}


\subsection{Experimental Setup}\label{sec:exp_setup}

In order to demonstrate the efficacy of the proposed MpO methodology, we conducted three design experiments: 1) a configurable packet parser \ifthenelse{\equal{\Hidden}{Yes}}{[hidden for blind review]}{\cite{SantiagodaSilva:2018}}; 2) a flow-based traffic manager (TM) \cite{Benacer:2018}; and, 3) a digital up-converter \cite{xilinx_up}.

The first design experiment is a configurable packet parser briefly introduced in \S~\ref{sec:case_study}. 
To enable reproducibility, the code of this experiment is open-source \cite{cpp:parser}. 

The second design experiment is a flow-based TM architecture proposed by Benacer \textit{et al.} \cite{Benacer:2018} in the context of SDN. 
The architecture is made up of a traffic policer, a packet scheduler, a systolic priority queue, and a traffic shaper. 
This source code is proprietary.

The third design experiment is a digital up-converter retrieved from an application note from Xilinx \cite{xilinx_up}. 
The up-converter design is composed of multi-stage FIR filters, a direct digital synthesizer, and a mixer. 
This implementation is open-source \cite{cpp:duc}.

All experiments targeted a Xilinx Virtex-7 FPGA, part number XC7VX690TFFG1761-2.
Vivado HLS 2015.4 was used to generate synthesizable RTL code.
While we have tested more recent versions of Vivado HLS, according to our experiments, the version 2015.4 is the one that better supports modern C++ constructs.
Xilinx Vivado 2015.4 was used for the synthesis and place and route (P\&R). 
 Code complexity is presented in terms of equivalent lines of code (eLOC). The eLOC metric ignores blank and commented lines. LOC, when presented, represents the actual number of lines of code.
We measure code reuse with CCFinderX \cite{ccfindeer}, an open-source tool based on the work by Kamiya \textit{et al.}~\cite{Kamiya:2002}, to detect code clones.

\subsection{Results}\label{sec:experiments}


\subsubsection{Configurable Packet Parser}\label{sec:parser_res}


\begin{table}[]
\centering
\ifthenelse{\equal{\Hidden}{No}}{
\caption{Packet parser results. Adapted from \cite{SantiagodaSilva:2018}}
}{
\caption{Packet parser results. Adapted from [hidden for blind review]}
}

\label{tab:parser_resutls}
\small{
\begin{tabular}{|c|S[table-format=3.1]|S[table-format=2.1]|S[table-format=5.0]|S[table-format=5.0]|S[table-format=5.0]|}
\hline
\textbf{Work} & \textbf{\begin{tabular}[c]{@{}c@{}}Freq.\\ {[}MHz{]}\end{tabular}} &  \textbf{\begin{tabular}[c]{@{}c@{}}Lat.\\ {[}ns{]}\end{tabular}} & \textbf{LUTs} & \textbf{FFs} & \textbf{Slices}  \\
\hline
VHDL~\cite{Benacek:2016}                 & 195.3    & 27              & N/A            & N/A           & 8000    \\ 
\cite{Benacek:2016}                      & 195.3    & 46.1            & 10103          & 5537          & 15640   \\ 
\cite{Benacek:2016} MpO                  & 312.5    & 41.6            & 6450           & 10308         & 16758    \\
MpO                                      & 312.5    & 25.6            & 6046           & 8900          & 14946    \\ 
\hline
\end{tabular}}
\end{table}

Table~\ref{tab:parser_resutls} presents the results of the configurable packet parser experiment. In terms of throughput (omitted from the table) and latency, this work performs as well as the hand-crafted VHDL implementation reported in \cite{Benacek:2016}. 
This work outperforms automatically generated VHDL code in all aspects except in the number of FFs.
 The LUTs reduction can be explained by the degree of parameterization that our specialized C++ classes offer. 
The operations are therefore fine-tuned for each header instance.

We have conducted a different experiment where we mimic Benacek's architecture using our MpO methodology. 
This experiment is labelled ``\cite{Benacek:2016} MpO" in Table~\ref{tab:parser_resutls}. 
Architectural aspects aside, this hybrid implementation delivers better results than the original Benacek implementation, significantly reducing the latency (-10\%) and the number of LUTs (-35\%). 
One takeaway from this experiment is that VHDL lacks in abstraction to be used as a direct conversion language from a high-abstraction DSL, such as P4. 
On the other hand, C++ offers an adequate dialect to represent network semantics that can be described using P4.

\subsubsection{Flow-based traffic manager}\label{sec:tm_results}

Table~\ref{tab:tm_results} presents the results of the TM implementation. This TM implementation can process 1024 different packet flows. The queue depth of this TM is 128.
To provide a fair comparison for this experiment, we did not perform any algorithmic or architectural optimization in the original code. Also, we kept the same optimization directives of the original design.

\begin{table}[]
\centering
\caption{Flow-based TM results}
\label{tab:tm_results}
\small{
\begin{tabular}{|c|S[table-format=3.1]|S[table-format=6.0]|S[table-format=6.0]|S[table-format=5.0]|S[table-format=4.0]|}
\hline

\textbf{Work} & \textbf{\begin{tabular}[c]{@{}c@{}}Freq.\\ {[}MHz{]}\end{tabular}} & \textbf{LUTs} & \textbf{FFs} & \textbf{Slices} &\textbf{eLOC}\\
\hline
\multicolumn{6}{|c|}{\textbf{Systolic Slice Size = 3}} \\
\hline
\cite{Benacer:2018}  & 91.5                & 37581          & 13723           & 9833   & 784\\ 
\cite{Benacer:2018} MpO  & 102.4               & 33575           & 13536          & 9182   & 1001\\ 
\hline
\multicolumn{6}{|c|}{\textbf{Systolic Slice Size = 4}} \\
\hline
\cite{Benacer:2018} MpO  & 74.5                 & 55625           & 13891          & 14669 & 1001\\ 
\hline
\multicolumn{6}{|c|}{\textbf{Systolic Slice Size = 8}} \\
\hline
\cite{Benacer:2018} MpO  & 44.9                & 116666           & 13585          & 31884 & 1001\\ 
\hline
\multicolumn{6}{|c|}{\textbf{Systolic Slice Size = 16}} \\
\hline
\cite{Benacer:2018} MpO  & 31.0                & 200450           & 13930          & 57876 & 1001\\ 
\hline
\end{tabular}}
\end{table}

Besides code modernization using C++11 constructs, we augmented the degree of parameterization of the TM design.
The core component of this TM is a systolic implementation of a priority queue. In the original design, each systolic slice implemented a micro queue of two or three elements.
The MpO implementation fully parameterizes these micro queues, not limiting to two or three. This can be seen in Table~\ref{tab:tm_results}, in which we show the TM implementation results, with 4, 8, and 16 elements in each systolic slice.

The MpO version of the TM improved HW QoR. Noticeably, the circuit frequency was improved by more than 10\%. The area consumption was improved as well, with a reduction of more than 10\% in LUTs and 7\% in occupied slices. No effects in latency, II, DSPs, and BRAMs were observed.

The MpO implementation augmented eLOC by 27\%. Indeed, this was expected because we generalized a hardwired implementation of the systolic queue slice to support arbitrary systolic slices. Moreover, a significant contributor to the increased eLOC is a library that can be reused elsewhere. This library has roughly 10\% of the total eLOC, in which we implemented type trait classes and generic helper functions. In both original and MpO-based implementation, CCFinderX did not find code clones. 

\subsubsection{Digital up-converter}\label{sec:up_res}

Table~\ref{tab:duc_results} shows the results of the digital up-converter implementation. We did not perform optimizations on the original code. We only modified the code for the FIR filters. While HW QoR results consider the whole design, the SW quality analysis applies only to the filters.

\begin{table}[]
\centering
\caption{Digital up-converter HW QoR results}
\label{tab:duc_results}
\small{
\begin{tabular}{|c|S[table-format=3.1]|c|S[table-format=4.0]|S[table-format=4.0]|S[table-format=4.0]|}
\hline
\textbf{Work} & \textbf{\begin{tabular}[c]{@{}c@{}}Freq.\\ {[}MHz{]}\end{tabular}} & \textbf{\begin{tabular}[c]{@{}c@{}}Lat.\\ {[}cycles{]}\end{tabular}} & \textbf{LUTs} & \textbf{FFs} & \textbf{Slices} \\
\hline
\cite{xilinx_up}  & 371.6    & 3394$\sim$3395            & 3472           & 7388          & 1641 \\ 
\cite{xilinx_up} MpO  & 404.0    & 3375$\sim$3376            & 3010           & 5723          & 1568 \\ 
\hline
\end{tabular}}
\end{table}

As shown in Table~\ref{tab:duc_results}, the MpO approach improves QoR metrics compared to the original digital up-converter implementation from Xilinx. There were improvements in the maximum frequency, latency, and area consumption, notably for FFs. 
The FFs reduction can be explained by the reduced latency, which means that a shorter pipeline was required in the MpO implementation. No effects on BRAM, DSP, and II were observed, thus, not reported in Table~\ref{tab:duc_results}.

The MpO approach significantly improves software quality as presented in Table~\ref{tab:duc_results_sw}.
The measure of eLOC in Table~\ref{tab:duc_results_sw} shows how expressive the MpO is compared to the original design. 
The MpO-based code is 16\% more concise than its original counterpart. 
Also, we evaluate code reuse by measuring code clone patterns as reported in Table~\ref{tab:duc_results_sw}.
We observe that CCFinderX found 6 patterns of code clones in the original design while no clones were found in our implementation.
Indeed, the MpO methodology favors code reuse and STL usage, following a DRY methodology to avoid code duplication.
In addition, in the original design, CCFinderX found an average of 2.33 replicated instances per clone pattern, with a maximum of 3. CCFinderX also reported an average of 83.5 LOC per clone, with a maximum of 115.

\begin{table}[]
\centering
\caption{Digital up-converter SW quality results}
\label{tab:duc_results_sw}
\small{
\begin{tabular}{|c|S[table-format=3.0]|S[table-format=1.0]|c|c|}
\hline
 \textbf{Work} & \textbf{eLOC} & \textbf{Clones} & \(\scriptstyle  \mathrm{\overline{\frac{Instances}{Clone}}}\) & \(\scriptstyle \mathrm{\overline{\frac{LOC}{Clone}}}\) \\ 
\hline
Original \cite{xilinx_up}   & 383 & 6 & 2.33  & 83.5\\ 
\cite{xilinx_up} MpO   & 323 & 0 & $\varnothing$ & $\varnothing$ \\ 
\hline
\end{tabular}}
\end{table}

\subsection{Analysis and Discussions}\label{sec:discussions}


Zhao and Hoe \cite{Zhao:17} present quantitative results for the design of a network-on-chip. The authors compare their methodology to an auto-generated RTL implementation, while comparisons to hand-crafted RTL are not shown. On average, their results show an overhead of 11\% and 8\% for the LUTs consumption and the clock period. FF usage is reduced by 58\%. Latency results are not presented. Their experiment is similar to the comparison between this work applied to the packet parser and \cite{Benacek:2016}. Using our methodology, the maximum frequency is $1.6\times$ higher, the latency is reduced by 45\%, and LUTs by 40\%, while increasing the number of FFs by 60\%. These improvements in the LUTs consumption and the maximum frequency are due to our design's ability to specialize operations, leading to faster and more compact circuits.

Oezkan \textit{et al.} \cite{Oezkan:17} present comparative results between their HLS-based image processing library and the results of an image processing DSL that generates C++ code. In that comparison, their results outperform the auto-generated code, which is expected, since their library is directly hand-crafted in C++. Therefore, their results cannot be used as a baseline for a fair comparison against our proposed methodology.

While similar works have exploited modern C++ with HLS design \cite{Muck:14,Muck:14_,Thomas:2016,Richmond:2018}, no generalized methodology has been presented to date. Muck and Frohlich \cite{Muck:14,Muck:14_} have focused on unified CPU-FPGA C++ code-base. Thomas \cite{Thomas:2016} and Richmond \textit{et al.} \cite{Richmond:2018} have exploited the power of modern C++ to implement features not natively supported by HLS tools, such as recursion and high-order functions. None of these works have presented the benefits of using C++ in the generated HW, as we have shown. Also, no SW quality metrics have been presented in these works.

\section{Conclusion}\label{sec:conclusion}

HLS is a game changer to spread FPGA usage outside the HW world. However, achieving high QoR with HLS design still relies on detailed FPGA knowledge to generate FPGA-friendly low-level code, an uncommon skill for software developers. Such codes lower the design abstraction level making their comprehension and maintenance tedious even for experienced programmers. This HLS design approach follows a uni-dimensional design perspective, trading-off HW QoR and SW quality.


In this work, we introduced a bi-dimensional HLS design view by proposing the MpO methodology. The MpO methodology targets FPGA development with HLS exploiting standard C++11 constructs. The proposed MpO methodology builds on the concept of an MpO base class. The five presented characteristics of an MpO base class leverage HLS design, improving HW QoR, code readability and modularity while raising the abstraction development level. Through three design examples, we showed that using the MpO methodology, a C++ code can deliver results comparable to hand-crafted VHDL design. We as well showed that the code complexity can be reduced using the zero-overhead characteristic of C++. 


%% file: bare_conf.bbl
\begin{thebibliography}{10}
\providecommand{\url}[1]{#1}
\csname url@samestyle\endcsname
\providecommand{\newblock}{\relax}
\providecommand{\bibinfo}[2]{#2}
\providecommand{\BIBentrySTDinterwordspacing}{\spaceskip=0pt\relax}
\providecommand{\BIBentryALTinterwordstretchfactor}{4}
\providecommand{\BIBentryALTinterwordspacing}{\spaceskip=\fontdimen2\font plus
\BIBentryALTinterwordstretchfactor\fontdimen3\font minus
  \fontdimen4\font\relax}
\providecommand{\BIBforeignlanguage}[2]{{%
\expandafter\ifx\csname l@#1\endcsname\relax
\typeout{** WARNING: IEEEtran.bst: No hyphenation pattern has been}%
\typeout{** loaded for the language `#1'. Using the pattern for}%
\typeout{** the default language instead.}%
\else
\language=\csname l@#1\endcsname
\fi
#2}}
\providecommand{\BIBdecl}{\relax}
\BIBdecl

\bibitem{Matai:14}
\BIBentryALTinterwordspacing
J.~Matai \emph{et~al.}, ``{Enabling FPGAs for the Masses},'' \emph{CoRR}, vol.
  abs/1408.5870, 2014. [Online]. Available:
  \url{http://arxiv.org/abs/1408.5870}
\BIBentrySTDinterwordspacing

\bibitem{Winterstein:13}
F.~Winterstein \emph{et~al.}, ``{High-level synthesis of dynamic data
  structures: A case study using Vivado HLS},'' in \emph{2013 International
  Conference on Field-Programmable Technology (FPT)}, Dec 2013, pp. 362--365.

\bibitem{Muck:14}
T.~R. Muck and A.~A. Frohlich, ``{Toward Unified Design of Hardware and
  Software Components Using C++},'' \emph{IEEE Transactions on Computers},
  vol.~63, no.~11, pp. 2880--2893, Nov 2014.

\bibitem{Muck:14_}
\BIBentryALTinterwordspacing
------, ``{"A metaprogrammed C++ framework for hardware/software component
  integration and communication"},'' \emph{Journal of Systems Architecture},
  vol.~60, no.~10, pp. 816 -- 827, 2014. [Online]. Available:
  \url{http://www.sciencedirect.com/science/article/pii/S1383762114001143}
\BIBentrySTDinterwordspacing

\bibitem{hoare:1972}
C.~Hoare, ``{The quality of software},'' \emph{Software: Practice and
  Experience}, vol.~2, no.~2, pp. 103--105, 1972.

\bibitem{stroustrup:2015}
B.~Stroustrup and H.~Sutter, ``{C++ Core Guidelines},''
  \url{http://isocpp.github.io/CppCoreGuidelines/CppCoreGuidelines}, 2018.

\bibitem{Cong:2011}
J.~Cong \emph{et~al.}, ``{High-Level Synthesis for FPGAs: From Prototyping to
  Deployment},'' \emph{IEEE Transactions on Computer-Aided Design of Integrated
  Circuits and Systems}, vol.~30, no.~4, pp. 473--491, April 2011.

\bibitem{liang:12}
Y.~Liang \emph{et~al.}, ``{High-level synthesis: productivity, performance, and
  software constraints},'' \emph{Journal of Electrical and Computer
  Engineering}, vol. 2012, p.~1, 2012.

\bibitem{Homsirikamol:14}
E.~Homsirikamol and K.~Gaj, ``{Can high-level synthesis compete against a
  hand-written code in the cryptographic domain? A case study},'' in \emph{2014
  International Conference on ReConFigurable Computing and FPGAs (ReConFig14)},
  Dec 2014, pp. 1--8.

\bibitem{Liu:2016}
\BIBentryALTinterwordspacing
X.~Liu \emph{et~al.}, ``{High Level Synthesis of Complex Applications: An H.264
  Video Decoder},'' in \emph{Proceedings of the 2016 ACM/SIGDA International
  Symposium on Field-Programmable Gate Arrays}, ser. FPGA '16.\hskip 1em plus
  0.5em minus 0.4em\relax New York, NY, USA: ACM, 2016, pp. 224--233. [Online].
  Available: \url{http://doi.acm.org/10.1145/2847263.2847274}
\BIBentrySTDinterwordspacing

\bibitem{Zhou:2018}
\BIBentryALTinterwordspacing
Y.~Zhou \emph{et~al.}, ``{Rosetta: A Realistic High-Level Synthesis Benchmark
  Suite for Software Programmable FPGAs},'' in \emph{Proceedings of the 2018
  ACM/SIGDA International Symposium on Field-Programmable Gate Arrays}, ser.
  FPGA '18.\hskip 1em plus 0.5em minus 0.4em\relax New York, NY, USA: ACM,
  2018, pp. 269--278. [Online]. Available:
  \url{http://doi.acm.org/10.1145/3174243.3174255}
\BIBentrySTDinterwordspacing

\bibitem{Winterstein:14}
F.~Winterstein \emph{et~al.}, ``{Separation Logic-Assisted Code Transformations
  for Efficient High-Level Synthesis},'' in \emph{2014 IEEE 22nd Annual
  International Symposium on Field-Programmable Custom Computing Machines}, May
  2014, pp. 1--8.

\bibitem{Gao:2016}
\BIBentryALTinterwordspacing
X.~Gao \emph{et~al.}, ``{Automatically Optimizing the Latency, Area, and
  Accuracy of C Programs for High-Level Synthesis},'' in \emph{Proceedings of
  the 2016 ACM/SIGDA International Symposium on Field-Programmable Gate
  Arrays}, ser. FPGA '16.\hskip 1em plus 0.5em minus 0.4em\relax New York, NY,
  USA: ACM, 2016, pp. 234--243. [Online]. Available:
  \url{http://doi.acm.org/10.1145/2847263.2847282}
\BIBentrySTDinterwordspacing

\bibitem{Cong:17}
\BIBentryALTinterwordspacing
J.~Cong \emph{et~al.}, ``{Bandwidth Optimization Through On-Chip Memory
  Restructuring for HLS},'' in \emph{Proceedings of the 54th Annual Design
  Automation Conference 2017}, ser. DAC '17.\hskip 1em plus 0.5em minus
  0.4em\relax New York, NY, USA: ACM, 2017, pp. 43:1--43:6. [Online].
  Available: \url{http://doi.acm.org/10.1145/3061639.3062208}
\BIBentrySTDinterwordspacing

\bibitem{Coplien:1995}
\BIBentryALTinterwordspacing
J.~O. Coplien, ``{Curiously Recurring Template Patterns},'' \emph{C++ Rep.},
  vol.~7, no.~2, pp. 24--27, Feb. 1995. [Online]. Available:
  \url{http://dl.acm.org/citation.cfm?id=229227.229229}
\BIBentrySTDinterwordspacing

\bibitem{Thomas:2016}
D.~B. Thomas, ``{Synthesisable recursion for C++ HLS tools},'' in \emph{2016
  IEEE 27th International Conference on Application-specific Systems,
  Architectures and Processors (ASAP)}, July 2016, pp. 91--98.

\bibitem{Richmond:2018}
D.~Richmond, A.~Althoff, and R.~Kastner, ``{Synthesizable Higher-Order
  Functions for C++},'' \emph{IEEE Transactions on Computer-Aided Design of
  Integrated Circuits and Systems}, pp. 1--1, 2018.

\bibitem{Eran:19}
H.~Eran, L.~Zeno, Z.~Istvanz, and M.~Silberstein, ``{Design Patterns for Code
  Reuse in HLS Packet Processing Pipelines},'' in \emph{To appear at the 2019
  IEEE 27th Annual International Symposium on Field-Programmable Custom
  Computing Machines (FCCM)}, April 2019, pp. 1--10.

\bibitem{Zhao:17}
Z.~Zhao and J.~C. Hoe, ``{"Using Vivado-HLS for Structural Design: a NoC Case
  Study"},'' Carnegie Mellon University, ECE Department, Pittsburgh, PA USA,
  Tech. Rep., 2017.

\bibitem{Oezkan:17}
M.~A. Oezkan \emph{et~al.}, ``{A Highly Efficient and Comprehensive Image
  Processing Library for C++-based High-Level Synthesis},'' in \emph{FSP 2017;
  Fourth International Workshop on FPGAs for Software Programmers}, Sept 2017,
  pp. 1--10.

\bibitem{Kapre:16}
N.~Kapre and S.~Bayliss, ``{Survey of domain-specific languages for FPGA
  computing},'' in \emph{2016 26th International Conference on Field
  Programmable Logic and Applications (FPL)}, Aug 2016, pp. 1--12.

\bibitem{Bosshart:14}
\BIBentryALTinterwordspacing
P.~Bosshart \emph{et~al.}, ``{P4: Programming Protocol-independent Packet
  Processors},'' \emph{SIGCOMM Comput. Commun. Rev.}, vol.~44, no.~3, pp.
  87--95, Jul. 2014. [Online]. Available:
  \url{http://doi.acm.org/10.1145/2656877.2656890}
\BIBentrySTDinterwordspacing

\bibitem{Wang:2017}
\BIBentryALTinterwordspacing
H.~Wang \emph{et~al.}, ``{P4FPGA: A Rapid Prototyping Framework for P4},'' in
  \emph{Proceedings of the Symposium on SDN Research}, ser. SOSR '17.\hskip 1em
  plus 0.5em minus 0.4em\relax New York, NY, USA: ACM, 2017, pp. 122--135.
  [Online]. Available: \url{http://doi.acm.org/10.1145/3050220.3050234}
\BIBentrySTDinterwordspacing

\bibitem{Khan:2017}
J.~Khan and P.~Athanas, ``{Creating Custom Network Packet Processing Pipelines
  on HMC-Enabled FPGAs}.''

\bibitem{Sultana:2017}
\BIBentryALTinterwordspacing
N.~Sultana \emph{et~al.}, ``{Emu: Rapid Prototyping of Networking Services},''
  in \emph{2017 {USENIX} Annual Technical Conference ({USENIX} {ATC}
  17)}.\hskip 1em plus 0.5em minus 0.4em\relax Santa Clara, CA: {USENIX}
  Association, 2017, pp. 459--471. [Online]. Available:
  \url{https://www.usenix.org/conference/atc17/technical-sessions/presentation/sultana}
\BIBentrySTDinterwordspacing

\bibitem{Singh:2008}
S.~Singh and D.~J. Greaves, ``{Kiwi: Synthesis of FPGA Circuits from Parallel
  Programs},'' in \emph{2008 16th International Symposium on Field-Programmable
  Custom Computing Machines}, April 2008, pp. 3--12.

\bibitem{SantiagodaSilva:2018}
\BIBentryALTinterwordspacing
J.~Santiago~da Silva, F.-R. Boyer, and J.~M.~P. Langlois, ``{P4-Compatible
  High-Level Synthesis of Low Latency 100 Gb/s Streaming Packet Parsers in
  FPGAs},'' in \emph{Proceedings of the 2018 ACM/SIGDA International Symposium
  on Field-Programmable Gate Arrays}, ser. FPGA '18.\hskip 1em plus 0.5em minus
  0.4em\relax New York, NY, USA: ACM, 2018, pp. 147--152. [Online]. Available:
  \url{http://doi.acm.org/10.1145/3174243.3174270}
\BIBentrySTDinterwordspacing

\bibitem{Gibb:2013}
G.~Gibb \emph{et~al.}, ``Design principles for packet parsers,'' in
  \emph{Architectures for Networking and Communications Systems}, Oct 2013, pp.
  13--24.

\bibitem{cpp:constexpr}
{Jeferson Santiago da Silva}, ``{constexpr Example},''
  \url{https://godbolt.org/z/fpme-0}, 2019.

\bibitem{cpp:zero_abstraction}
------, ``{C++ Zero Abstraction Example},'' \url{https://godbolt.org/z/xkKXON},
  2019.

\bibitem{cpp:lambda}
------, ``{Lambda versus Functor Example},''
  \url{https://godbolt.org/g/uQqU65}, 2019.

\bibitem{Kim:2017}
J.~H. Kim \emph{et~al.}, ``{FPGA-based CNN inference accelerator synthesized
  from multi-threaded C software},'' in \emph{2017 30th IEEE International
  System-on-Chip Conference (SOCC)}, Sept 2017, pp. 268--273.

\bibitem{cpp:optim}
{Jeferson Santiago da Silva}, ``{Frontend Optimizations Example},''
  \url{https://godbolt.org/g/dheE2Q}, 2019.

\bibitem{Noronha:2017}
D.~H. Noronha \emph{et~al.}, ``{Rapid circuit-specific inlining tuning for FPGA
  high-level synthesis},'' in \emph{2017 International Conference on
  ReConFigurable Computing and FPGAs (ReConFig)}, Dec 2017, pp. 1--6.

\bibitem{Benacer:2018}
I.~Benacer \emph{et~al.}, ``{Design of a Low Latency 40 Gb/s Flow-Based Traffic
  Manager Using High-Level Synthesis},'' in \emph{2018 IEEE International
  Symposium on Circuits and Systems (ISCAS)}, May 2018, pp. 1--5.

\bibitem{xilinx_up}
A.~Paek and J.~Wu, ``{Designing a Digital Up-Converter using Modular C++
  Classes in Vivado High Level Synthesis},'' 2016.

\bibitem{cpp:parser}
{Jeferson Santiago da Silva}, ``{Packet Parser Code},''
  \url{https://github.com/engjefersonsantiago/P4HLS}, 2019.

\bibitem{cpp:duc}
------, ``{Digital-up Converter Code},''
  \url{https://github.com/engjefersonsantiago/MpO/tree/master/DUC}, 2019.

\bibitem{ccfindeer}
{Peter Senna Tschudin}, ``{CCFinderX Code},''
  \url{https://github.com/petersenna/ccfinderx-core}, 2019.

\bibitem{Kamiya:2002}
T.~Kamiya \emph{et~al.}, ``Ccfinder: a multilinguistic token-based code clone
  detection system for large scale source code,'' \emph{IEEE Transactions on
  Software Engineering}, vol.~28, no.~7, pp. 654--670, Jul 2002.

\bibitem{Benacek:2016}
P.~Ben\'{a}cek \emph{et~al.}, ``{P4-to-VHDL: Automatic Generation of 100 Gbps
  Packet Parsers},'' in \emph{2016 IEEE 24th Annual International Symposium on
  Field-Programmable Custom Computing Machines (FCCM)}, May 2016, pp. 148--155.

\end{thebibliography}
